\documentclass[onecolumn,showpacs,showkeys,preprintnumbers,amsmath,amssymb, 12 pt]{revtex4}

\usepackage{graphicx}

\usepackage{dcolumn}
\usepackage{bm}

\providecommand{\beqa}{\begin{eqnarray}}
 \providecommand{\bf}{\mathbf}
 \providecommand{\rm}{\mathrm}
\providecommand{\eeqa}{\end{eqnarray}}

\def\Z2{{\mathbf{Z}_2}}

\usepackage{amssymb}
\usepackage{url,hyperref}

\begin{document}


\title{First Test of High Frequency Gravity Waves from Inflation using ADVANCED LIGO }

\author{Alejandro Lopez}
\email{aolopez@umich.edu}

\author{Katherine Freese}
\email{ktfreese@umich.edu}

\affiliation{Michigan Center for Theoretical Physics, University of
Michigan, Ann Arbor, Michigan 48109-1040, USA}

\date{\today}

\begin{abstract}

Inflation models ending in a first order phase transition produce gravitational waves (GW) via bubble collisions of the true vacuum phase. We demonstrate that these bubble collisions can leave an observable signature in Advanced LIGO, an upcoming ground-based GW experiment. These GW  are dependent on two  parameters of the inflationary model: $\varepsilon$ represents the energy difference between the false vacuum and the true vacuum of the inflaton potential, and $\chi$ measures how fast the phase transition ends ($\chi \sim$ the number of e-folds during the actual phase transition). Advanced LIGO will be able to test the validity of single-phase transition models within the parameter space $10^7 \rm{GeV}\lesssim \varepsilon^{1/4} \lesssim 10^{10} \rm{GeV}$ and $0.19 \lesssim \chi \lesssim 1$.
If inflation occurred through a first order phase transition, then Advanced LIGO could be the first to discover high frequency GW from inflation.

\end{abstract}


\maketitle
\section*{Introduction}
Several experiments are underway to detect the stochastic
gravitational background from early universe cosmology. These
gravity waves (GWs) propagate almost freely throughout the entire
history of the universe and thus can be a direct source of
information about the Universe at very early times.  One  source of GWs is inflation \cite{Guth:1980zm}, a superluminal growth
phase of the Universe which can explain the homogeneity and isotropy of the Universe as well as
the generation of density perturbations required for structure formation.
There are two types of contributions to GW from inflation.  In slowly  rolling
models, quantum fluctuations of space-time lead to contributions proportional
to the height of the inflaton potential.   In tunneling models, where the inflation ends in a first order phase transition, there is an additional
 contribution due to bubble collisions of true vacuum bubbles at the end of the phase transition \cite{Turner:1992tz,Kosowsky:1992vn,Kosowsky:1992rz}. In this paper, we will present the possible gravitational wave signatures that Advanced LIGO could detect  originating from vacuum bubble collisions in inflationary models that end in a rapid tunneling event.   

There are two widely different origins of GW from inflationary cosmology.  The most commonly studied are those due to quantum fluctuations of the inflation field in slowly rolling models of inflation.  Whereas the scalar modes of the fluctuations produce density perturbations that seed structure formation, the tensor modes produce gravity waves.  These GW are on very large scale lengths, comparable to the present-day horizon scale.  Cosmic microwave background experiments can in principle search for these GW. The Wilkinson Microwave Anisotropy Probe (WMAP) and Planck satellites have not found GW yet but instead place bounds on their contribution to the energy density of the Universe. Specifically, WMAP bounds GW in the range of frequencies $10^{-18} \text{ Hz}<f<10^{-16}\text{ Hz}$ \cite{Komatsu:2008hk,Smith:2005mm,WMAP,WMAP9} at the level of roughly  $h^2\Omega_{GW} \lesssim 10^{-15}$, where $\Omega_{GW} $ is the fraction of the Universe's (critical) energy density in the form of GW.  The bounds from Planck are currently only slightly better but should improve once the polarization data is analyzed \cite{Planck}. Most recently, the BICEP2 experiment has claimed detection of gravitational waves at frequencies $f \sim 2 \times 10^{-17}$ Hz \cite{BICEP}. BICEP2 measured the tensor to scalar ratio with a mean value $r=0.2$ at a pivot scale $k_*=0.002$ Mpc$^{-1}$. A tensor to scalar ratio $r=0.2$ implies that $h^2\Omega_{GW}\cong 1.3 \times 10^{-15}$. At first glance, the results on the value of the tensor to scalar ratio from WMAP and Planck seem to contradict those of BICEP2. There have been many different proposals to alleviate the tension between WMAP/Planck and BICEP2, such as allowing for the running of the scalar spectral index \cite{nrun1, nrun2}, the inclusion of sterile neutrino contributions \cite{sterile1,sterile2}, and other considerations \cite{acceleration1,acceleration2}. The incorporation of dust polarization maps to the BICEP2 data could also explain the discrepancy between experiments \cite{dust1}. The polarization data from the Planck experiment, expected to be released in the near future, will be an important test in collaborating BICEP2's claim of detection.   
Proposed for the future is CMBPOL, which would have a sensitivity at least an order of magnitude better than existing CMB experiments \cite{Smith:2005mm}.

In this paper we study the second possible origin of GW from inflation, bubble collisions in tunneling models of inflation that end in a first order phase transition.  The physical signatures of these bubbles are on much smaller length scales. Adv LIGO can be used to test GW densities due to bubble collisions roughly in the range of $10\text{ Hz}<f<200\text{ Hz}$ and is sensitive up to $h^2\Omega_{GW}\sim 10^{-9}$. The current $95 \%$ upper bound of LIGO S5 is around $h^2\Omega_{GW} \sim 10^{-6}$ in the frequency band $41.5$-$169.25$ Hz. Consequently LIGO S5 has not been able to test any of the first order phase transitions studied in this paper \cite{WMAP}. The largest amplitude considered is approximately $h^2\Omega_{GW} \sim 10^{-8}$ as will be shown in the Results section. LIGO S6 has also gathered data, but its bounds are very similar to those of LIGO S5 and thus would also not be able to elucidate on the first order phase transition considered here. Proposed for
the more distant future are the space-interferometers BBO and DECIGO, with possible launch within $20$ or $30$ years \cite{Smith:2005mm}\cite{BBO-DECIGO}.    

We also wish to mention previous wok on bubble collisions in a {\it thermal} background; i.e. the bubbles nucleate in a radiation dominated background, rather than in the vacuum dominated background of inflationary transitions.   Thermal bubbles differ from vacuum bubbles in that they have extra structure such as turbulence which produce relevant signatures \cite{Kusenko1,Kusenko2,Giblin1,Hindmarsh1,Tinatin1,Tinatin2}. Another source of gravitational waves that has captured much interest in recent years is preheating, which also has an elaborate structure due to the thermal background \cite{Easther1, Easther2, Felder1, GarciaBellido1, Easther3, Dufaux1, GarciaBellido2, Price1, Dufaux2}.  While all of this work on bubble collisions in a thermal background is very interesting it does not apply to the vacuum bubbles from inflation studied here.

In this paper we restrict our studies to bubble collisions in inflationary models with one single tunneling event.  In future work we will turn to the possibility of multiple tunneling events such as seen in chain inflation \cite{Freese1,Freese2,Freese3,Freese4,Freese5,Chialva1}.  The purpose of this paper is to determine if Adv LIGO would be able to measure the GW energy spectrum of single phase transition tunneling models; and if so, which region of the parameter space they would be able to ``see".

\section*{Tunneling  inflationary models}

In tunneling models of inflation,  the universe starts out in a high-energy minimum and then tunnels down to its global minimum. During the time spent in the ``false" vacuum, the universe expands superluminally by  $\sim$ 60 (or so) $e$-folds required to resolve the flatness problem. There are two competing factors which constrain single-phase transition model building: the horizon/flatness problems on the one hand and percolation on the other. In order to explain the lack of intrinsic curvature of the universe, a Grand Unified (GUT) scale inflationary model needs the universe to inflate by approximately 60 $e$-folds. Thus, it is necessary for the universe to stay in the ``false" minimum long enough to expand $e^{60}$ times its original size. On the other hand, the phase transition needs to be rapid enough such that bubbles of true vacuum intersect one another and percolation is complete. Both criteria can be met if we introduce a time-dependent nucleation rate $\Gamma$, which gives the probability per physical volume per time that a bubble will be produced in a region still in the ``false" minimum. 

The failure of  ``old" inflation (Guth's (1981) original inflation model \cite{Guth:1980zm})  was due primarily to assuming a constant nucleation rate. Old inflation produced a `Swiss Cheese Universe':  to allow for sufficient inflation, the tunneling rate had to be so slow that the phase transition was never able to complete  \cite{GuthandWeinberg}.  At the end of inflation most of the Universe remained in a state of false vacuum but contained disconnected bubbles of true vacuum.  The bubbles of true vacuum were unable to merge together sufficiently to percolate and thermalize, as would be required for an end to inflation. 

As a  solution to this problem Adams and Freese (1991) as well as Linde (1990) suggested models with a time-dependent tunneling rate.  The tunneling rate starts out very slow, so that sufficient inflation can take place; and then suddenly the tunneling rate becomes very fast so that the phase transition quickly percolates and completes, allowing for a Universe consisting entirely of true vacuum.  The time dependence of the nucleation rate has been proposed by Adams and Freese (1991) and by Linde (1990) to arise from multi-field interactions \cite{linde,Adams:1991ma,Freese:2004vs}.  Equivalently (using different terminology) in a multi-dimensional potential, at first the field slowly rolls in one field direction for at least 60 e-folds of inflation, and after that tunnels rapidly in a different field direction to complete the phase transition.  Cort$\hat{e}$s and Liddle (2009) studied these models in light of WMAP data and concluded they are still viable \cite{Liddle}.

This paper applies to any inflationary model that ends in a first order phase transition.  Other examples besides the double-field model mentioned above include models with a scalar field non-minimally coupled to gravity \cite{Notari1}\cite{Notari2}.

 The nucleation rate of true vacuum bubbles in the sea of false vacuum is given by
\begin{equation}
\Gamma(t) = Ae^{-S(t)},
\end{equation} 
where $S(t)$ is the action for the bounce
solution extrapolating between false and true vacua
and A is a determinant constant with units of [Mass]$^4$ \cite{ColemanI} \cite{ColemanII}\cite{Turner:1992tz}\cite{Kosowsky:1992vn}. We follow the work of C. Caprini, R. Durrer and G. Servant (2007) and expand our action to first order around $t_*$, which will be defined below as the time when the universe became $99\%$ true vacuum. Thus, the nucleation rate can be expressed as 
\begin{equation}
\label{eq:betadef}
\Gamma(t)=\Gamma(t_*)e^{-\beta(t-t_*)}.
\end{equation} 
Here $\beta=\frac{dS(t)}{dt}|_{t_*}$ is a parameter whose inverse sets a rough  time scale for the phase transition to complete in our inflationary model (more accurately see Equation [9] below) \cite{Caprini:2007xq}.   

The probability of a point staying in the false vacuum, $p(t)$, can be calculated from the nucleation rate:
\begin{equation}\label{prob}
p(t)=e^{-I(t)},
\end{equation}
where $I(t)$ is given by
\begin{equation}\label{eq:FVexp}
I(t)=\int_{-\infty}^t dt' \Gamma(t') a^3(t')\frac{4\pi}{3} r(t,t')^3. 
\end{equation}
Here $a(t)$ is the scale factor of the Friedmann-Lemaitre-Robertson-Walker metric.  A bubble nucleated at time $t'$ grows to have a radius
$r(t,t')$ at a later time $t$ given by 
\begin{equation}
r(t,t')=\int_t'^t dt'' \frac{c}{a(t'')},
\end{equation}
where we have assumed that the wall of the bubbles expand at the speed of light $c$, and will take $c=1$ for all future calculations. If we further assume that the phase transition occurs fast enough that we can neglect the expansion of the universe, then Equation [\ref{eq:FVexp}] simplifies to
\begin{equation}\label{I}
I(t)=\int_{-\infty}^t dt' \Gamma(t') \frac{4\pi}{3} (t-t')^3=\frac{8\pi}{\beta^4}\Gamma(t).
\end{equation}  
This assumption is justified if the duration of the phase transition is less than a Hubble time interval, $H^{-1}$ \cite{Kosowsky:1992rz}. 

 We can proceed to calculate the duration of the phase transition from $p(t)$\cite{Turner:1992tz}\cite{Kosowsky:1992vn}\cite{Caprini:2007xq}. The ``beginning" of the phase transition $t_m$ will be defined as the time when $1\%$ of the universe is found to be in the true vacuum; and similarly, the ``end" of the phase transition $t_*$ is given by the time when the universe is $99\%$ in the true vacuum. We will choose $m$ and $M$ such that $p(t_m)=e^{-m}\cong 1$ and $p(t_*)=e^{-M}\cong 0$. In other words, $m$ and $M$ satisfy
\begin{eqnarray}
1-p(t_m)=1-e^{-m}=0.01 \indent \Rightarrow \indent m=0.01 \\
1-p(t_*)=1-e^{-M}=0.99 \indent \Rightarrow \indent M=5.0 .
\end{eqnarray}
 Thus, the duration of the phase transition, i.e. the time it takes for the universe to move from $1\%$ to $99\%$ true vacuum, is given by
\begin{equation}\label{duration}
t_*-t_m=\ln(\frac{M}{m})\beta^{-1}.
\end{equation} 
For the values of $m$ and $M$ chosen above, 
\begin{equation}
\label{eq:deltat}
t_*-t_m \sim 6 \beta^{-1} \, .
\end{equation}

 It should be noted that this time duration is not a measurement of how long the universe stayed in the ``false" vacuum. The exponential nature of the nucleation rate allows the ``false" minimum to be stable for a long time, providing the necessary expansion that resolves the flatness problem; and proceeds to rapidly destabilize the ``false" vacuum so that it can tunnel quickly and percolation is achieved. Thus, the duration of the phase transition as defined by Equation [\ref{duration}] measures the pace at which vacuum changes from being $1\%$ to $99\%$ in the global minimum.  In the model where the Universe slowly rolls for a long time before changing direction in field space and rapidly tunneling, this is the time for the tunneling only.
   
It will serve convenient to take advantage of the natural cosmological time parameter, the Hubble time, to define
\begin{equation}\label{chidef}
\chi = \ln\left(\frac{M}{m}\right) H \beta^{-1} \cong 6H\beta^{-1}.
\end{equation}
One can roughly think of $\chi$ as the number of e-foldings during the tunneling transition, as evidenced by Equation [\ref{chidef}] and Equation [\ref{duration}].
Therefore, the assumption of having a phase transition faster than a Hubble time interval constrains the values of $\chi$ that can be consistently studied. We require that
\begin{equation}
H(t_*-t_m)=H\ln(\frac{M}{m})\beta^{-1} < 1 
\end{equation}
\begin{equation}
\Rightarrow \indent \chi < 1.
\end{equation}
Thus, the single-phase transition inflationary models studied here will be restrained to have 
\begin{equation}
\label{eq:nohubble}
\chi \le 1 \, .
\end{equation}  
With the choices leading to Eqn. (\ref{eq:deltat}), this constraint amounts to no more than 1 e-fold during the tunneling (as expected since the origin of this constraint is that the simplified equations we are using only apply if we can neglect the expansion of the Universe.)

 In addition any physically viable inflationary model needs to percolate. We will follow the arguments made by Turner et al.(1992) in order to utilize their  bound on the nucleation rate to further constrain the parameter $\chi$ \cite{Turner:1992tz}. In order for the bubbles to ``outrun" the general cosmic expansion, a successful inflationary transition needs to decrease the actual physical volume in the false vacuum, $V_{phys} \propto a^3(t)p(t)$. Nucleated bubbles will be able to ``outrun" the inflationary expansion once 
 \begin{equation}
 V_{phys}^{-1}\frac{dV_{phys}}{dt}=3H-\frac{dI}{dt} <0 \, .
\end{equation}  
 The ability to decrease the physical volume found in the false vacuum does not guarantee percolation but is a necessary condition. We define ``$t_{e}$" to be the time at which this criterion is satisfied, $\frac{dI}{dt}|_{t_e}=3H$ (i.e. $\frac{dV}{dt}|_{t_e}=0$). Turner et al. (1992) found a lower bound on a time-dependent nucleation rate
\begin{equation}
\label{eq:percolate}
\frac{\Gamma}{H^4}\Big|_{t_e} > \frac{9}{4\pi},
\end{equation}
which we apply here.  We note that Guth and Weinberg (1983) had originally found a lower bound on a {\it constant} nucleation rate in order to achieve percolation, applicable to the case of old inflation (which failed exactly because it does not percolate)\cite{GuthandWeinberg}.  The work of Turner et al (1992), on the other hand, is for time-dependent nucleation rates as relevant here. 

 We would like to convert the constraint in Eqn. (\ref{eq:percolate})  to a bound on $\chi$. Utilizing Equation (\ref{I}) and taking $t=t_e$, we find that
\begin{eqnarray}
\frac{dI}{dt}\Big|_{t_e}=3H=8\pi \left(\frac{\chi}{\ln{\frac{M}{m}}}\right)^3 H \frac{\Gamma}{H^4}\Big|_{t_e} \\
\Rightarrow \frac{\Gamma}{H^4}\Big|_{t_e}=\frac{3}{8\pi}\left(\frac{\ln{\frac{M}{m}}}{\chi}\right)^3
\end{eqnarray}
Thus the bound in Eq. (\ref{eq:percolate})  implies the following upper bound on $\chi$:
\begin{equation}
\frac{\Gamma}{H^4}\Big|_{t_e}=\frac{3}{8\pi}\left(\frac{\ln{\frac{M}{m}}}{\chi}\right)^3 > \frac{9}{4\pi} \iff \chi < \left(\frac{1}{6}\right)^{1/3}\ln{\left(\frac{M}{m}\right)}  \sim 3.4.
\end{equation}

It should be noted that this bound is less restrictive than the one in Eqn.(\ref{eq:nohubble}); any model satisfying $\chi\le 1$ automatically satisfies the percolation bound. We require both constraints to hold, i.e., the inflationary transition goes from $1\% - 99 \%$ true vacuum sufficiently fast to ignore the expansion of the universe and the physical volume of false vacuum decreases, allowing for the Universe to percolate.

\subsection*{Gravity Waves from a Single First Order Phase Transition}

 The spectrum of gravity waves (GWs) from multi-bubble collisions at a tunneling phase transition (PT)
 with energy difference $\varepsilon$ between false and true vacua was
worked out numerically in \cite{Kosowsky:1992rz} and subsequently more accurately by
\cite{Huber} and \cite{Caprini:2007xq}. The parameters $\varepsilon$ and $\chi$ will be the only two free variables that characterize the spectrum of single PT models.  In this paper we will follow the results of Huber and Konstandin (2008)  \cite{Huber}. They computed the GW spectrum resulting from multi-bubble collisions produced by a time-dependent exponential nucleation rate. Their simulations assumed a phase transition that lasted much less than $H^{-1}$ and thus justified neglecting the expansion of the universe. The assumptions taken by Huber et al. (2008) are the same taken in the previous section, which enable us to take full advantage of their results. Here we examine  which region of the parameter space $\{\chi, \varepsilon\}$ Advanced LIGO would  be able to observe.   

 The gravitational wave energy spectrum is defined to be
\begin{equation}
\Omega_{GW}(f)=\frac{1}{\rho_c}\frac{d\rho_{GW}}{d\ln(f)},
\end{equation}  
where $f$ is the frequency, $\rho_c$ is the current critical density and $\rho_{GW}$ is the gravitational wave energy density.
The GW energy spectrum depends on the wall velocity of the bubbles ($v_b$), the fraction of vacuum energy to radiation energy ($\alpha=\frac{\rho_{vac}}{\rho_{rad}}$), and the efficiency at which the vacuum energy is transformed into kinetic energy of the bulk fluid instead of reheating the plasma inside the bubble ($\kappa$) \cite{Huber}. For purposes of this paper, we will take the wall velocity to be close to the speed of light( $v_b\cong 1$), assume the strong-detonation limit ($\alpha \rightarrow \infty$), and take $\kappa =1$ so that the vacuum energy is converted almost to its entirety into the kinetic (rather than thermal) energy of the bubble. This further supports our assumption of taking the wall velocity to be close to the speed of light, since the constant pressure felt by the bubble wall will force it to accelerate to relativistic speeds very quickly \cite{Kosowsky:1992vn}. Henceforth  we will take the results found by Huber and Konstandin (2008) in \cite{Huber} and take the limits: $v_b=1$, $\alpha \rightarrow \infty$ and $\kappa=1$. These are the limits necessary to study vacuum bubbles \cite{Kosowsky:1992vn}.  

The numerical results of Huber and Konstandin (2008) \cite{Huber} can roughly be fit in the following three frequency regimes as 
\begin{equation} \label{GWprofile}
h^2\Omega_{GW}(f) \cong
\left\{
	\begin{array}{ll}
		f^3 & \mbox{} f \ll f^{peak}  \\
		h^2\Omega^{peak}\frac{3.8 (\frac{f}{f^{peak}})^{2.8}}{1+2.8(\frac{f}{f^{peak}})^{3.8}} & \mbox{ } f \approx f^{peak}
		\\
		f^{-1} & \mbox{} f \gg f^{peak}
	\end{array}
\right. 
\end{equation}
Here $\Omega^{peak}$ is the GW energy spectrum evaluated at the peak frequency and will be defined in terms of physical parameters of the phase transition shortly. Here $h$ is the current Hubble parameter in units of $100$ km/sec/Mpc.   In order to work with an analytic expression for the spectrum, we construct the GW energy spectrum as a piece-wise function in the following manner

\begin{equation} \label{GWprofile0}
h^2\Omega_{GW}(f) \cong
\left\{
	\begin{array}{ll}
		h^2\Omega^{peak}(\frac{f}{F})^3  & \mbox{if } f < \eta f^{peak} \hspace{2 mm} \textrm{with} \hspace{2 mm} \eta<1 \\
		h^2\Omega^{peak}\frac{3.8 (\frac{f}{f^{peak}})^{2.8}}{1+2.8(\frac{f}{f^{peak}})^{3.8}} & \mbox{if } f \geq \eta f^{peak}\hspace{2 mm} \textrm{with} \hspace{2 mm} \eta<1.
	\end{array}
\right. 
\end{equation}
We determine the value of the constants $\eta$ and $F$ by demanding the spectrum to be sufficiently smooth, i.e., we take the matching conditions 
\begin{eqnarray}\label{alpha}
\lim_{f\rightarrow \eta f^{peak -}} h^2\Omega_{GW}(f)=\lim_{f\rightarrow \eta f^{peak +}} h^2\Omega_{GW}(f) \\
\lim_{f\rightarrow \eta f^{peak -}} \frac{d h^2\Omega_{GW}(f)}{d\ln f}\cong \lim_{f\rightarrow \eta f^{peak +}}\frac{d h^2\Omega_{GW}(f)}{d\ln f}.
\end{eqnarray} 
In Eq.(24) we accept an error of $7\%$. Then we find that the GW energy spectrum has the following approximate structure
\begin{equation}\label{GWprofile}
h^2\Omega_{GW}(f) \cong
\left\{
	\begin{array}{ll}
		 h^2\Omega^{peak} 38(\frac{ f}{f^{peak}})^3  & \mbox{if } f < 10^{-5} f^{peak} \\
		h^2\Omega^{peak}\frac{3.8 (\frac{f}{f^{peak}})^{2.8}}{1+2.8(\frac{f}{f^{peak}})^{3.8}} & \mbox{if } f \geq 10^{-5} f^{peak}.
	\end{array}
\right. 
\end{equation}
This enables us to estimate the numerical calculations for the GW energy spectrum found by Huber and Konstandin (2008) in an analytic fashion. 

The peak frequency of the GW spectrum is determined by the characteristic timescale of the phase transition, $\beta^{-1}$ (see Eq.(\ref{eq:betadef})). Specifically, Huber et al. (2008) find for the peak frequency
\begin{equation}\label{fpeak} 
f^{peak}=0.23 \beta \, .
\end{equation}
They also find
\begin{equation} \label{Omegapeak}
\hspace{10 mm} h^2\Omega^{peak}=h^2\frac{\varepsilon }{\rho_c}\kappa^2\left(\frac{H}{\beta}\right)^2\left(\frac{\alpha}{\alpha+1}\right)^2\left(\frac{0.11v_b^3}{0.42+v_b^2}\right)\cong \frac{0.002 h^2}{\rho_c}\varepsilon \chi^2 \, 
\end{equation}
for the peak amplitude, where we have taken  $v_b=1$, $\kappa=1$ and $\alpha \rightarrow \infty$ in the final equality.  Redshifting  the frequency and energy density of gravitational radiation
as $a^{-1}$ and $a^{-4}$ respectively, we find that at the current epoch (subscript $0$):
\begin{eqnarray}\label{f^0-omega0}
f^{peak}_{0}= f^{peak}(\frac{a_*}{a_0}),\\
h^2\Omega^{peak}_0=h^2\Omega^{peak}(\frac{a_*}{a_0})^4 \label{f^0-omega01},
\end{eqnarray}
where the subscript ``*" denotes the time $t_*$ at which the phase transition ended. Assuming that reheating is instantaneous allows us to evaluate $\frac{a_*}{a_0} \cong 7.6 \times 10^{-14} \left(\frac{100}{g_*}\right)^{1/3}\left(\frac{1 \text{ GeV}}{T_*}\right)$. $T_*$ is the temperature increase right after a single phase transition,
\begin{equation}\label{temp}
T_*=\left(\frac{30\varepsilon}{g_*\pi^2}\right)^{1/4};
\end{equation} 
and the total number of relativistic degrees of freedom at temperature $T_*$ is taken to be $g_* \cong 100$. However, the process of reheating could be fairly complicated. Depending on the specific model considered, the reheating epoch could also last for some time. The details of the reheating epoch will depend on how the kinetic energy of the walls is converted into heat. Reheating in the context of a first order phase transition has been considered in \cite{Kolb, Zhang}. Furthermore, it was shown by R. Watkins and L. Widrow that bubble collisions convert the energy in the bubble walls efficiently into scalar radiation \cite{Watkins}. Nevertheless, the duration of the reheating epoch will depend on the details of the phase transition. Precise numerical studies would be needed to study a specific reheating model, and understand the duration and equation of state of the Universe during the reheating epoch. In the Appendix we consider the effects of a non-instantaneous reheating epoch on the gravitational wave energy spectrum. With the exception of the Appendix, we will assume for all future calculations that reheating was instantaneous. 

This assumption simplifies the calculation, and allows us to calculate the GW energy spectrum as a function of only two parameters: $\varepsilon$ and $\chi$. Combining Equations [\ref{fpeak}, \ref{Omegapeak}] and [\ref{temp}], the dependence of $f_0^{peak}$ and $h^2\Omega_0^{peak}$ on $\{  \varepsilon, \chi \}$ becomes clear. The peak frequency and GW energy density per critical density are given by
\begin{equation}\label{fpeak0}
f^{peak}_0=9.35 \times 10^{-8} \frac{\varepsilon^{1/4}}{1\mbox{GeV}}\frac{1}{\chi}  \mbox{Hz}
\end{equation}
\begin{equation}\label{Omegapeak0}
h^2\Omega^{peak}_0(\chi)=5.9 \times 10^{-8} h^2 \chi^2. 
\end{equation}
Using these two equations in Eqn. [\ref{GWprofile0}] evaluated at the current epoch, we obtain the GW energy spectrum expected today for different choices of the two parameters $\{ \varepsilon, \chi \}$. We will study which range of parameter space can be measured by Adv. LIGO.

\section*{Results}

The two parameters $\{\chi,\varepsilon\}$ in tunneling inflation determine the GW. 
The peak frequency  in  Eq. [\ref{fpeak}]  depends on both parameters, while $\Omega_{GW}$
in Eq. [\ref{Omegapeak}] depends on $\chi$ (but not on $\varepsilon$). Since the GW energy density scales as $\chi^2$, the largest GW amplitude is found
for the largest value of $\chi$ studied in the paper, the value $\chi =1$ allowed by the bound in Eqn.[\ref{eq:nohubble}].

 The GW energy spectrum for tunneling models is plotted in Figure [1] for frequencies ranging 
 from $10^{-6}$ Hz to $10^{8}$ Hz.  Adv. LIGO is in the middle of this frequency range, approximately around $10 {\rm Hz} <f<100$ Hz \cite{WMAP}. In Figure [1], $\chi$ is kept constant at a value of $\chi=1$, and  the value of $\varepsilon^{1/4} \in \{10^4,10^{8.5},10^{11}\}$ is varied. In making this plot we have chosen the highest value of  $\chi$ treated in this paper since it produces the largest observable signal with  $h^2\Omega_0^{peak} \propto \chi^2$.
 For parameters $\chi=1$ and $\varepsilon^{1/4}=10^{8.5}$
the GW are observable in Advanced LIGO which is
sensitive to stochastic
signals with $\Omega_{\rm GW}h^2\gtrsim 10^{-9}$. 

 One can investigate the dependence of the GW on the value of $\chi$.
 Figure [2] shows a plot of the GW energy spectrum per critical density with $\varepsilon=10^{8.5}$ and varying values of $\chi$. The frequency range plotted in Figure [2] is from $0.1$ Hz to $10^{5}$ Hz in order to study more carefully the dependence of $\chi$ in the GW energy spectrum. For smaller values of $\chi$, the peak frequency shifts to higher values whereas the amplitude of gravitational waves decreases, as evident by Equations [\ref{fpeak} and \ref{Omegapeak}] and Figure[2].


\begin{figure}[t]
\includegraphics[height=130mm,width=180mm]{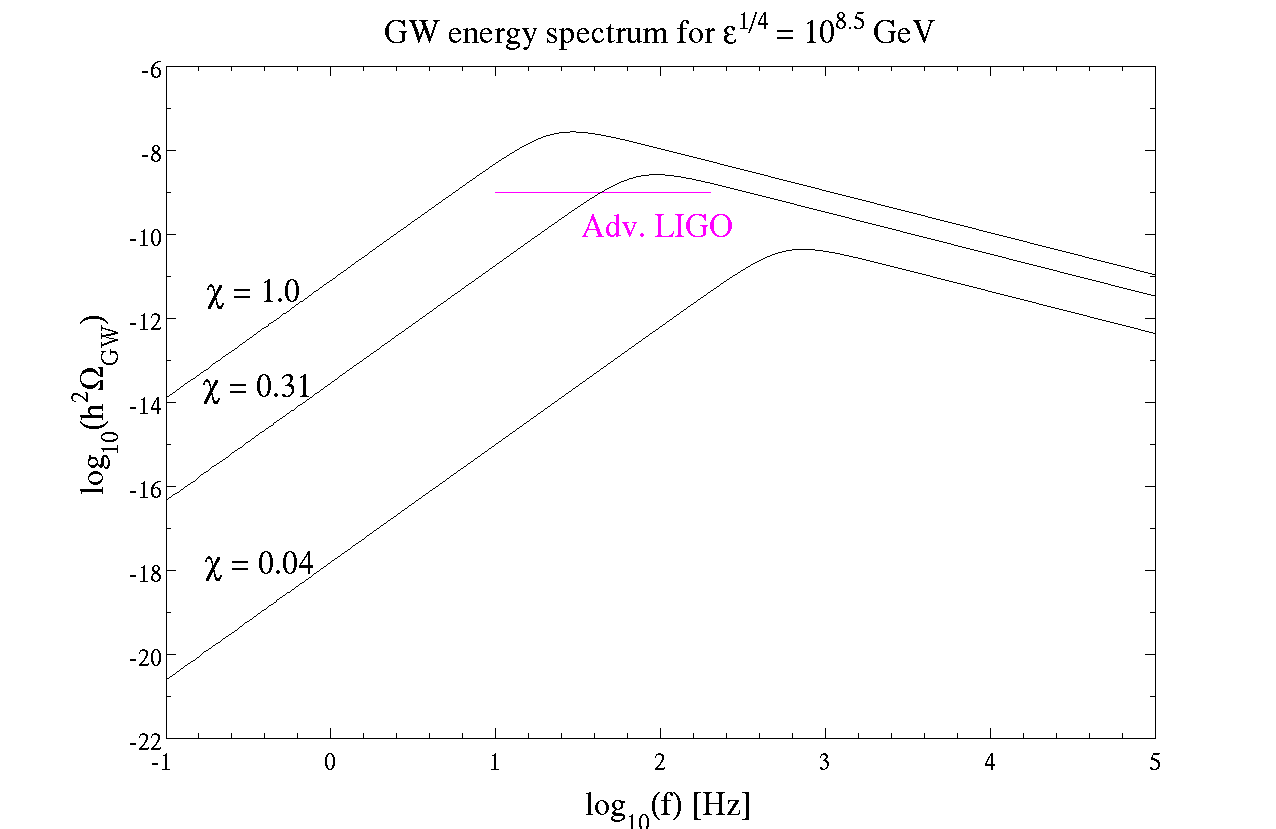}
 \caption{The spectrum of GWs produced from bubble collision in a tunneling phase transition for $\chi=1$ and various values of $\varepsilon$ (the energy difference between vacua). Different values of $\varepsilon$ only shift the peak frequency of the spectrum, but do not alter the amplitude of $\Omega_{GW}$. Here $\chi$ is the number of e-folds during the tunneling transition (not the same as the total number of inflationary e-folds).  The expected reach of Advanced LIGO is indicated by the horizontal line. }\label{bubble-var-e0}
\end{figure}

\begin{figure}[t]
\includegraphics[height=130mm,width=180mm]{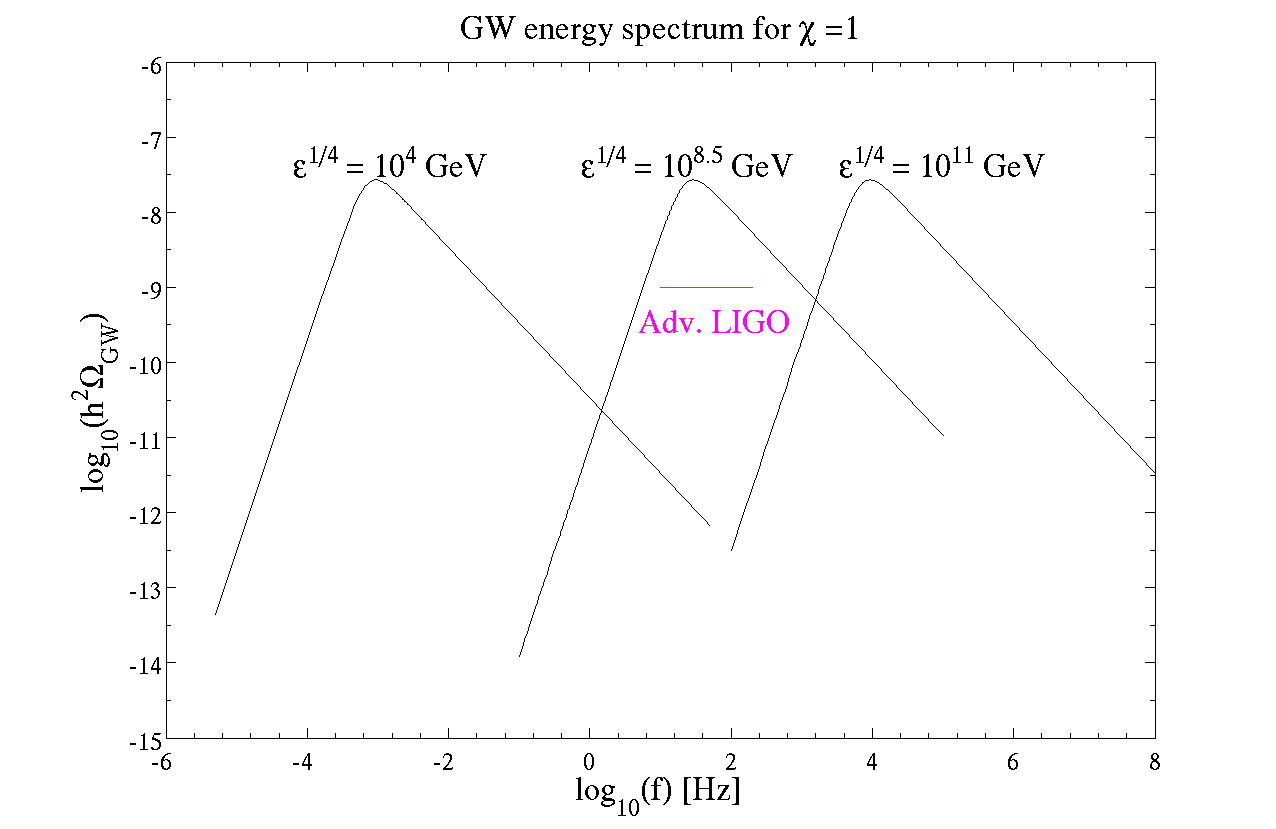}
 \caption{ GW energy spectrum produced in a tunneling inflationary model with a difference in energy between vacua of $\varepsilon^{1/4}=10^{8.5}$ and varying values of $\chi$.  Parameters are as described in Figure 1.  The expected reach of Advanced LIGO is indicated by the horizontal line. One can see that, for
smaller values of $\chi$, the peak frequency shifts to higher
values and the amplitude of gravitational waves decreases.}
\label{bubble-var-e}
\end{figure}


When we vary the value of $\varepsilon$, the GW amplitude remains the same while the spectrum shifts to a different frequency range.  For $ 10^{7}$ GeV $\lesssim \varepsilon^{1/4} \lesssim 10^{10}$
GeV, at least a part of the spectrum from
bubble collisions falls into the Advanced
LIGO sensitivity region.

\begin{figure}[h]
\includegraphics[height=115mm,width=180mm]{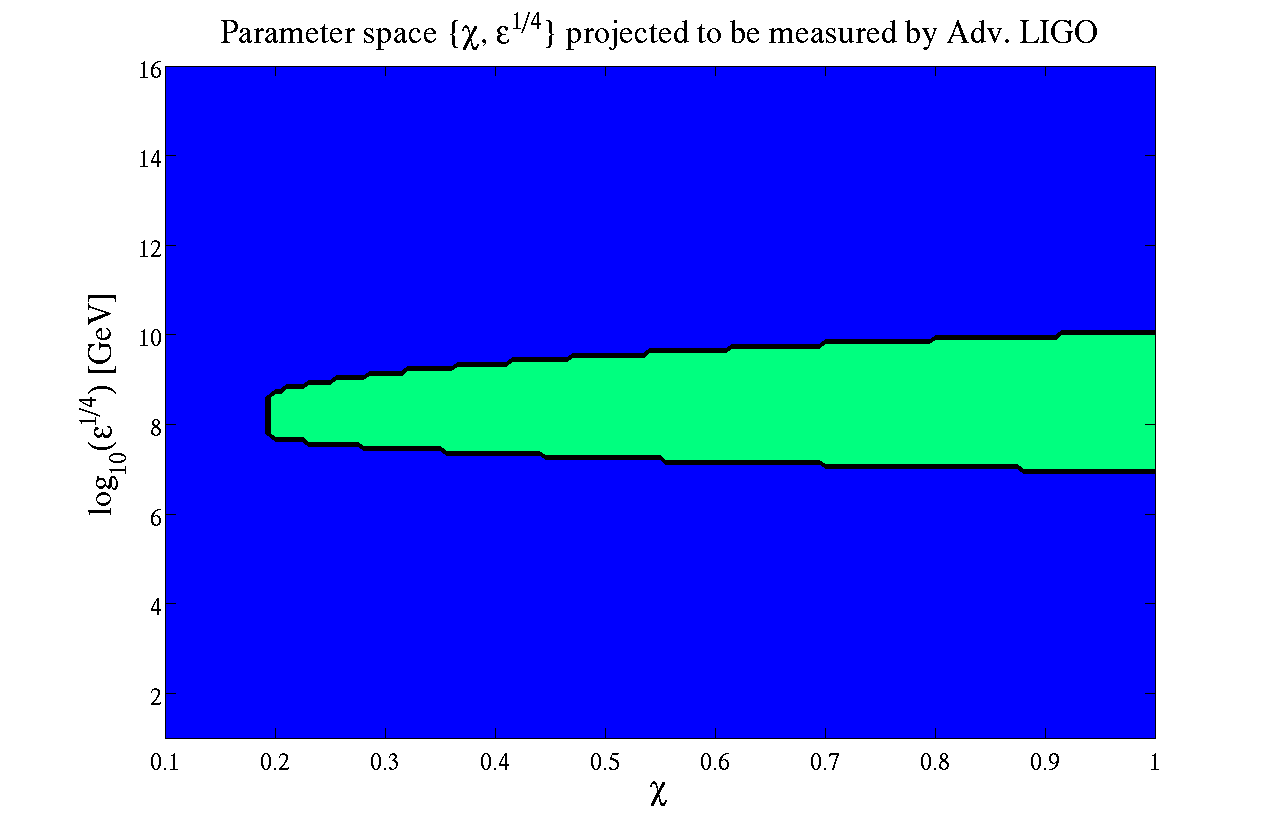}
 \caption{ This figure shows the range of $\{\chi,\varepsilon^{1/4}\}$ in single-tunneling transition inflationary models that could potentially be probed by Adv. LIGO (in green). Parameters are as described in Figure 1.  The range of $\chi$ goes up to $\chi=1$, because that is the highest value allowed by our assumption of having a PT which lasts less than a hubble time interval.}\label{chi}
\end{figure}


\section*{Conclusion}

We have shown in the previous section that Adv. LIGO will be able to measure and probe the GW energy spectrum  produced by  tunneling inflationary models. The parameter space of single PT inflationary models that can be measured in Adv. LIGO is shown in Figure [3]. Specifically, models with values of $0.19 \lesssim \chi \lesssim 1$ (where $\chi$ is the number of e-folds during the actual tunneling event) and $10^7 \rm{GeV} \lesssim \varepsilon^{1/4} \lesssim 10^{10} \rm{GeV}$ (where $\varepsilon$ is the energy difference between vacua) could potentially be tested by Adv. LIGO.  A positive signal could be a positive evidence in favor of single PT inflationary models, while the lack of a signal would potentially rule out the range of these models. 

The details of the reheating epoch right after the phase transition ended could also decrease the detectability of the stochastic GW signal. The calculations and figures shown in the Results section assume instantaneous reheating after the end of the phase transition. In contrast, if the Universe had an equation of state similar to matter domination ($w=0$) during the reheating phase, then the the peak frequency and GW energy density decrease by $(a_*/a_R)^{1/4}$ and $(a_*/a_R)$ respectively; where the subscript $``*"$ denotes the end of the phase transition, and $``R"$ specifies the time the Universe reaches thermal equilibrium and reheating ends. This result is shown in the Appendix. Given the projected sensitivity of Adv. LIGO, then there is a lower limit to the duration of the reheating phase $(a_*/a_R) \geq 0.036$ so that the GW signal can be detected. The decrease in the peak frequency and GW energy density could be fairly substantial and undetectable by Adv. LIGO in the case that the reheating epoch had an equation of state $w=0$. We refer the reader to the Appendix for the details of the computation. Nevertheless, it was shown in \cite{Watkins} that reheating after a first order phase transition could be very efficient. 

It is also an interesting question to ponder the details of the inflationary model prior to the tunneling that ends the inflation.
One possibility is that the inflaton field may be rolling down a nearly flat potential constrained by recent Planck measurements \cite{Planck}.  In many models the height of the potential would be near the GUT scale, not the case considered here. However, other models such as the Kinney-Mahanthappa version of natural inflation \cite{Kinney:1995ki} allow arbitrary potential heights.  Such rolling fields with $\epsilon$ in the right range could later tunnel to produce the signatures discussed in this paper.

Other proposed GW experiments, such as BBO and DECIGO, would be able to contribute further to the search for inflationary GW signals. These instruments could probe smaller frequencies with higher sensitivities and therefore study inflation models with smaller energy difference between vacua and faster phase transitions (i.e. lower values of $\varepsilon$ and $\chi$)\cite{BBO-DECIGO}.  

 In the future a theoretical study of GW produced in bubble collisions arising from slower phase transitions would be interesting, as these would also be testable by Advanced LIGO. Unfortunately this case is much more difficult.  In this paper we assumed that the phase transition lasted less than a Hubble time interval, i.e., we only considered values of $\chi\le 1$. In principle the value could be as high as $\chi<3.4$ and still percolate.  To go to higher values of $\chi$, the expansion of the Universe will have to be taken into account in the dynamics of the bubbles, rendering both analytic and numerical studies more complicated.  Nevertheless, since the GW amplitude scales as $\chi^2$, higher values of $\chi$ should be observable in Advanced LIGO; thus a study of this theoretically more difficult case is warranted in the future.  We also plan a future study of GW produced in chain inflation \cite{Freese1,Freese2,Freese3,Freese4,Freese5,Chialva1}, where the Universe tunnels through a series of phase transitions rather than merely one.
 
\section*{Acknowledgments}
We are grateful to Jim Liu for useful discussions. We thank Keith Riles for sharing GW detector sensitivity curves.  Research is
partially supported by NSERC of Canada.  This work was also supported in
part by the DOE under grant DOE-FG02-95ER40899 and by the Michigan
Center for Theoretical Physics.  K.F. was supported in part by a Simons Foundation Fellowship in Theoretical Physics.

\newpage
\appendix*
\section{Gravitational Wave signal after Reheating}

In this section we will consider the effects of having a non-instantaneous reheating epoch after the first order phase transition on the gravitational wave energy spectrum. We begin by rewriting Eqn. [\ref{f^0-omega0} and \ref{f^0-omega01}] as
\begin{eqnarray}\label{reheating_fpeak}
f^{peak}_{0}= f^{peak}\left(\frac{a_*}{a_R}\right)\left(\frac{a_R}{a_0}\right),\\
h^2\Omega^{peak}_0=h^2\Omega^{peak}\left(\frac{a_*}{a_R}\right)^4\left(\frac{a_R}{a_0}\right)^4 \label{reheating_omegapeak},
\end{eqnarray}
where $a_* = a(t_*)$ is the scale factor at the end of the phase transition, $a_R=a(t_R)$ is the scale factor at the end of reheating and $a_0$ is the scale factor today. Distinct reheating models will give different values for $(\frac{a_*}{a_R})$, since the duration and equation of state of the Universe depends on the details of the model. 
 
 During the reheating phase the homogeneous inflaton decays into lighter particles that will ultimately thermalize and acquire a black body spectrum at a temperature $T_R$. Once the Universe reaches thermal equilibrium, then the Hubble parameter at that time is given by $H_R^2=(8\pi/3)(\rho_R/M_{pl}^2)$,
where $M_{pl}=1.22 \times 10^{19}$ GeV is the Planck mass, $\rho_R=(g_R\pi^2/30)T_R^4$, and $g_R$ is the total number of relativistic degrees of freedom at temperature $T_R$. In future calculations we will approximate $g_R \cong 100$, although it could be a factor of 10 lower if we consider reheating to end at the beginning of Big Bang Nucleosynthesis. 

The duration and expansion history of the reheating phase, before the Universe reaches thermal equilibrium, is sensitive to the details and parameters of the reheating model. The reheating epoch has been studied in the context of first order phase transitions by  \cite{Watkins, Kolb, Zhang}.In particular, the Universe could have differing values for the equation of state during the reheating phase: $w=0$ similar to a matter dominated Universe or $w=1/3$ like in a radiation dominated Universe, among others. Thus, we write the Hubble parameter at the time of thermal equilibrium as
\begin{equation}\label{Hm}
H_R=H_* \left(\frac{a_*}{a_R}\right)^{\frac{3(1+w)}{2}};
\end{equation}      
where $w$ is the equation of state of the Universe during the reheating epoch. 

Furthermore, the scale factor at temperature $T_R$ is given by
\begin{eqnarray}\label{a*}
\frac{a_R}{a_0}=\left(\frac{g_0}{g_R}\right)^{1/3}\frac{T_0}{T_R}= \left(\frac{g_0^{1/3}}{g_R^{1/12}}\right) \left(\frac{8\pi^3}{90}\right)^{1/4}\frac{T_0}{\sqrt{H_RM_{pl}}}.
\end{eqnarray}
    
Combining Eq[\ref{reheating_fpeak} and \ref{reheating_omegapeak}] with equations [\ref{a*} and \ref{Hm}], the dependence of $f^{peak}_0$ and $h^2\Omega^{peak}_0$ on $\{\varepsilon$, $\chi, w\}$ becomes clear. The peak frequency and GW energy density per critical density today are given by
\begin{equation}\label{fpeak0}
f^{peak}_0=9.35 \times 10^{-8} \frac{\varepsilon^{1/4}}{1\mbox{GeV}}\frac{1}{\chi} \left(\frac{a_*}{a_R}\right)^{\frac{1-3w}{4}} \mbox{Hz}
\end{equation}
\begin{equation}\label{Omegapeak0}
h^2\Omega^{peak}_0(\chi)=5.9 \times 10^{-8} h^2 \chi^2 \left(\frac{a_*}{a_R}\right)^{1-3w}
\end{equation}

Note that the factor $\left(\frac{a_*}{a_R}\right)^{1-3w}$, which depends on the reheating parameters, equals to unity if $w=1/3$. In other words, the peak frequency and GW energy density measured today are unaffected by the details of the reheating model if the Universe had an equation of state similar to radiation domination right after the phase transition ended. In contrast, if the Universe had an equation of state $w=0$ during the reheating epoch, then the peak frequency and GW energy density decrease by $(a_*/a_R)^{1/4}$ and $(a_*/a_R)$ respectively. Specifically, if we take the overly conservative bound that reheating lasted until the beginning of Big Bang Nucleosynthesis and $w=0$, then the GW energy density decrease by a factor of ($10^{-13} - 10^{-17}$). This would make detection at Adv. LIGO impossible. Nevertheless, we would like to emphasize that this is an extreme case. The work of R. Watkins and L. Widrow suggest that reheating after a first order phase transition could be very efficient; comparable to slow-roll inflationary models.  

\end{document}